\documentclass[aps,prl,reprint,showpacs,amsmath,amssymb]{revtex4-1}
\usepackage{graphicx}

\begin{document}
\title{Impurity Scattering in Luttinger Liquid with Electron-Phonon Coupling}

\author{Alexey Galda, Igor V.\ Yurkevich and Igor V.\ Lerner}
\affiliation{School of Physics and Astronomy, University of
Birmingham, Birmingham B15 2TT, United Kingdom}

\date{\today}
\begin{abstract}
We study the influence of electron-phonon coupling on  electron transport through a Luttinger liquid with an embedded weak scatterer or weak link. We derive the renormalization group (RG) equations which indicate that the directions of RG flows can change upon  varying either the relative strength of the electron-electron and electron-phonon coupling or the ratio of Fermi to sound velocities. This results in the rich phase diagram with up to three fixed points: an unstable one with a finite value of conductance and two stable ones,
corresponding to an ideal metal or insulator.
\end{abstract}

\maketitle
\newcommand\ii{\mathrm{i}}
\newcommand\ee{\mathrm{e}}
\newcommand\dd{\mathrm{d}}
\newcommand\vF{v_{_{\rm F}}}
\newcommand\bF{\beta_{_{\rm F}}}

Interacting electrons in one dimension are known to form a Luttinger liquid (LL) characterized by power-law correlation functions \cite{Tom:50,*Lutt:63,*HALDANE:81,*vDSh:98}. This characteristic feature of the LL has been established via conductance measurements and a scanning tunneling microscopy (STM) both in carbon nanotubes \cite{Bockrath:99,*Yao:99,*Ishii:03,*Lee:04} and semiconductor quantum wires \cite{Auslaender:02,*Levy:06}.   Embedding a potential impurity  into the LL leads \cite{KaneFis:92a,*KF:92b,MatYueGlaz:93} to a universal (i.e.\ impurity-independent) power-law  (in temperature $T$) suppression of the transmission amplitude through the LL and suppression of the  tunneling density of states (TDoS)  near the impurity, with the latter fading away with the distance \cite{Eggert:00,GYL:04,*YL:05}.

The electron-phonon (el-ph) coupling in addition to the (Coulomb) electron-electron (el-el) repulsion in the LL is known to result in the formation of two polaron branches with different propagation velocities (see, for example,  \cite{Loss94,*F-BmixLutt}).
In the present Letter we show that embedding a single scatterer into such an el-ph liquid results in a rather rich phase diagram: depending on the relative strength of the el-el and el-ph coupling and on the ratio of the Fermi to sound velocities, the system can be
an ideal metal  (with conductance $g=1$ in the units of $e^2/h$ as in the absence of impurities \cite{MasStone:95}) or an ideal insulator for any impurity strength,  or to be in an intermediate state from which it can flow either to the metallic or insulating limit, depending on the impurity strength.

To show this we consider both the weak back-scatterer (WS) and weak link (WL) limit, the latter corresponding to a weak tunneling coupling between the two halves of the LL. In the phononless case the WS and WL amplitudes are known to scale at low energies as $\varepsilon ^{\widetilde{\gamma}_-}$ and $\varepsilon ^{\widetilde{\gamma}_+}$, respectively. For the repulsive el-el interaction the exponents $\widetilde{\gamma}_-\equiv K-1<0$ and $\widetilde{\gamma}_+=K^{-1}-1>0$ where $K<1$ is the Luttinger parameter. Therefore, the renormalization group (RG) flows always go in the direction of stronger scattering (weaker transmission) resulting at $T=0$ in the insulating phase for any scatterer \cite{KaneFis:92a}. The el-ph coupling leads to $\widetilde{\gamma}_\pm\to\gamma_\pm$, with each of the exponents $\gamma_\pm$ changing sign at different values of bulk parameters. In the WS limit which has been previously considered \cite{martin05}, the change of sign of $\gamma_-$ was reversing the RG flows
indicating that the weak scatterer becomes irrelevant for the sufficiently strong el-ph interaction and the LL thus remains in the metallic state. But the considerations of both the WS and WL limits presented here show that the change of signs in the exponents $\gamma_\pm$ happens at different values of the parameters, resulting in the rich phase diagram described in detail later.

To get the results, we employ the functional bosonization formalism in form developed in \cite{GYL:04}. This allows us first to include the  el-ph interaction which leads to electrons dressing with phonons, i.e.\  the formation of polarons, and only after that to bosonize the action. Prior to considering the embedded scatterer we describe  the polaron formation in a pure LL, reproducing the known results \cite{Loss94} in form convenient for further considerations.

We consider a model of 1D acoustic phonons linearly coupled to the electron density. The phonon spectrum  is assumed to be linear with a cutoff at the Debye frequency, $\omega_{_{\rm D}}=cq_{_{\rm D}}$ (a straightforward modification for the 3D or optical phonons will be described elsewhere). Integrating out the phonon field in the standard way  results in  substituting the dynamical coupling,
$    V({\xi })=V_0+\mathcal{D} (\xi ),
$ for the screened Coulomb interaction $V_0$ in the  LL action:
\begin{align}
{S}_{\rm LL}&= \!\! \!\sum_{\eta=\pm1} \int\!\!\dd \xi\,{\bar\psi}_{\eta}({\xi })\, \ii \partial_{\eta} \psi_{\eta}({\xi })-\frac{1}{2}  \int\!\!\dd \xi \,V ({\xi })\,n^2({\xi })\,.\label{LL}
\end{align}
Here $\partial_{\eta}\equiv \partial_{t}+\eta \vF \partial_{x}$,  the (spinless)  electron field is decoupled into the sum of left- ({$\eta\!=\!-1$}) and right- ({$\eta\!=\!1$}) moving terms, $\psi({\xi })=\psi_\mathrm{R}({\xi })\ee ^{\ii p_{_{\rm F}}x}+\psi_\mathrm{L}({\xi })\ee ^{-\ii p_{_{\rm F}}x}$ with $\xi \equiv ({x,t})$, and $n\equiv ({{\bar\psi}_\mathrm{R}\psi_\mathrm{R}+ {\bar\psi}_\mathrm{L}\psi_\mathrm{L}})$. We use the Keldysh formalism \cite{RS:86,*LevchKam} implying the time integration along the Keldysh contour in Eq.~(\ref{LL}) and below. The free phonon propagator $\mathcal{D}({\xi })$  in $    V({\xi })=V_0+\mathcal{D} (\xi ),
$  can be defined by the Fourier transform of its retarded component,
\begin{align}\label{D0}
\mathcal{D}_0^r({\omega,\,q})& =\frac{ \nu_0^{-1}\alpha_{\rm ph}\, \omega_q^2} { \omega_+ ^2-\omega_q^2}\,,&\omega_q&=cq\,,&\omega_+\equiv\omega+\ii0\,,
\end{align}
where  $c$ is the sound velocity, $\alpha_{\rm ph}$ is the dimensionless el-ph coupling constant, and  $\nu_0=(\pi \vF )^{-1}$ is the free spinless electron DoS.  The form of the LL action in Eq.~(\ref{LL}) implies neglecting the electronic backscattering. This remains justified in the presence of the el-ph coupling for low temperatures, $T\ll\omega_{_{\rm D}}$.

The next step is the Hubbard-Stratonovich transformation which decouples the $n^2$ term in the action (\ref{LL}) and results in the mixed fermionic-bosonic action in terms of the auxiliary bosonic field $\varphi$ minimally coupled to $\psi$:
\begin{equation}\label{S}
S_{\rm eff}=-\frac{1}{2}\int\!\!\dd \xi\,\varphi\,V^{-1}\,\varphi +\ii \!\! \int\!\!\dd \xi\, {\bar\psi}_{\eta}\left(\partial_{\eta}-\varphi\right)\psi_{\eta}\,.
\end{equation}
We  gauge out the coupling term by the transformation
 \begin{align}\label{gauge}
\psi_{\eta}({\xi }) &\to\psi_{\eta}({\xi }) \,\ee^{\ii\theta_{\eta}({\xi }) }\,,& \ii\partial_{\eta} \theta({\xi }) &=\varphi ({\xi })\,.
\end{align}
The Jacobian of this transformation results \cite{GYL:04} in substituting $  V^{-1}+\Pi$ for $V^{-1}$ in  Eq.~(\ref{S}), where $\Pi=\Pi_\mathrm{R} +\Pi_\mathrm{L}$ is the  one-loop electronic polarization operator (exact for the LL \cite{DzyalLar:73}), with $\Pi_\eta(\xi )=\ii g_\eta({\xi })g_\eta({-\xi }) $ and the free electron Green function defined via the Fourier transform of its retarded component as $g_\eta^r(\omega,\,q)=\left[\omega_+-\eta \vF  q\right]^{-1}$.

The phase  $\theta_{\eta}$ is related to the auxiliary field $\varphi$ by \cite{noteGaYL1}
\begin{equation*}
\theta_{\eta}(\xi )=\int \!\!\dd\xi '\,g^{B}_{\eta}(\xi ,\xi ')\,\varphi(\xi ')\,,
\end{equation*}
where $g^B_{\eta}$  is the bosonic Green functions that resolves Eq.~(\ref{gauge}). Its retarded component coincides with the free fermionic $g_{\eta}^r$ (while the Keldysh components  are naturally different).

The Green function of the interacting polarons is not gauge-invariant with respect to the transformation (\ref{gauge}) and depends on the correlation function $\ii U_{\eta\eta'}=\langle\theta_{\eta}\theta_{\eta'}\rangle$:
\begin{equation*}
G_{\eta}(\xi , \xi ')=g_{\eta}(\xi , \xi ')\,\ee^{\ii U_{\eta\eta}(\xi ,\xi ')}\,.
\end{equation*}
 The retarded Fourier component of $U_{\eta, \eta'}$ is found as
\begin{equation}\label{U}
U_{\eta\eta'}^r(\omega, q)= \frac{\omega_+\!+\!\eta'\vF q}{\omega_+\!-\!\eta\phantom' \vF q}\,
\frac{V_0 (\omega_+^2-\omega^2_q) +\nu_0^{-1}\alpha_\mathrm{ph}\omega^2_q} {\left(\omega_+^2-v^2_+q^2\right)\left(\omega_+^2-v^2_-q^2\right)},
\end{equation}
where $v_{\pm}$ are  velocities of the composite bosonic modes:
\begin{equation}\label{pm}
v^2_{\pm}=\frac{1}{2}\left[v^2+c^2\pm\sqrt{\left(v^2-c^2\right)^2 +4\alpha  \,v^2c^2}\,\right]\,.
\end{equation}
Here $v$ is the speed of plasmonic excitations in the phononless  LL,
$   v =\vF ({1+\nu_0V_0})^{1/2}\equiv \vF K^{-1},
$ where $K $ is the standard Luttinger parameter. Without phonons ($c=0$), one has $v_-=0$ and    $v_+ = v$, so that in this case (as well as for $\omega>\omega_{_{\rm D}}$  or $\alpha_\mathrm{ph}=0$), $U_{\eta \eta'} $ reduces to the usual LL plasmonic propagator.

We assume the parameter $\alpha\equiv\alpha_{\rm ph}K^2$ in Eq.~(\ref{pm})  obeying the inequality $\alpha<1$ to avoid the Wentzel--Bardeen instability \cite{Wentzel,*Bardeen:51} corresponding to   $v^2_- < 0$ (with the threshold shifted  from $\alpha_\mathrm{ph}=1$ for a pure el-ph model to $\alpha_\mathrm{ph}=1/K^2>1$). Then the velocities $v_\mp$ of the slow and fast composite bosonic modes in Eq.~(\ref{pm}) obey the inequalities $v_- < c,\,v < v_+$. These modes mean that the LL in the presence of the el-ph coupling becomes two-component, with the effective Luttinger parameters
\begin{equation*}
K_{\rm fast}={\vF }/{v_+}<K<1\,,\quad K_{\rm slow}={\vF }/{v_-}>1,
\end{equation*}
corresponding to el-el repulsion (which becomes stronger with the el-ph coupling) and attraction. It is the existence of these two modes that leads to a rich phase diagram when a scatterer is embedded.

Following Kane and Fisher \cite{KaneFis:92a} we consider two types of scatterers (assuming them pinned {at $x=0$} and not involved in lattice vibrations): a weak backscatterer (WS) or a weak tunneling link (WL). The WS action is
\begin{align}\label{WS}
S_{\rm ws}=\lambda_0\!\!\int\!\!\dd t\,{\bar \psi_\mathrm{R}(t)} \psi_\mathrm{L}(t)+{\rm c.c.},\;\psi (t)\equiv \psi(x\!=\!0,t)\,,
\end{align}
with $\lambda_0$ being a bare backscattering amplitude.

The WL action contains the tunneling term linking the two halves of the LL, labeled by $1$ and $2$:
\begin{eqnarray}\label{WL}
S_{\rm wl}=t_0\!\!\int\!\!\dd t\,{\bar\psi}_1 \psi_2+{\rm c.c.}, \quad\psi_{a}=\psi_{a\mathrm{R}}+\psi_{a\mathrm{L}}\,,
\end{eqnarray}
with $t_0$ being a bare tunneling amplitude. We begin with the WS case, noticing in advance that the WL case is dual to it as well as for the standard LL \cite{KF:92b}.

\begin{figure*}[t] \parbox{\columnwidth}{ \includegraphics[width=.85\columnwidth,height=0.6\columnwidth]{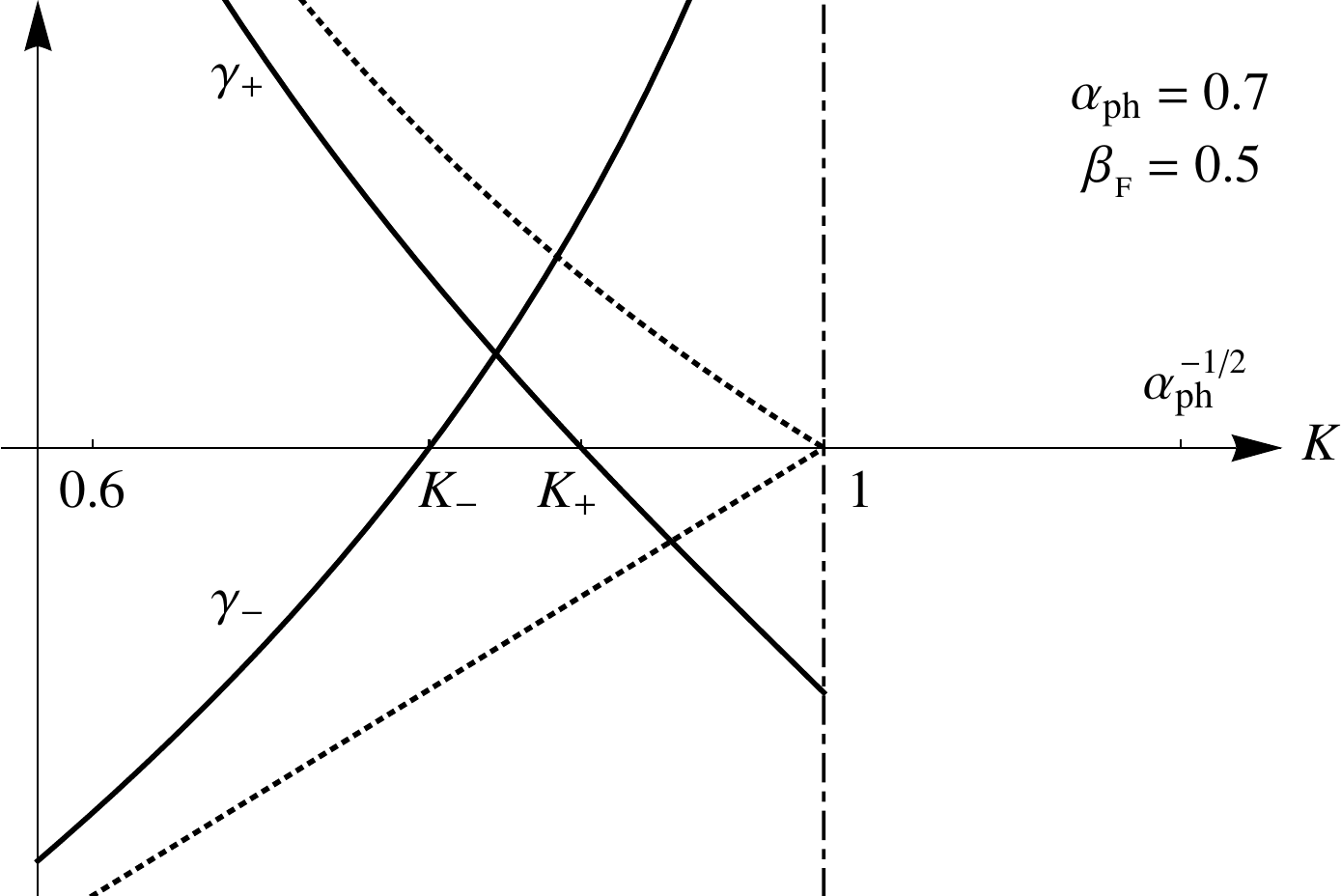}\\[9pt](a)} \parbox{\columnwidth}{ \includegraphics[width=.85\columnwidth,height=0.6\columnwidth]{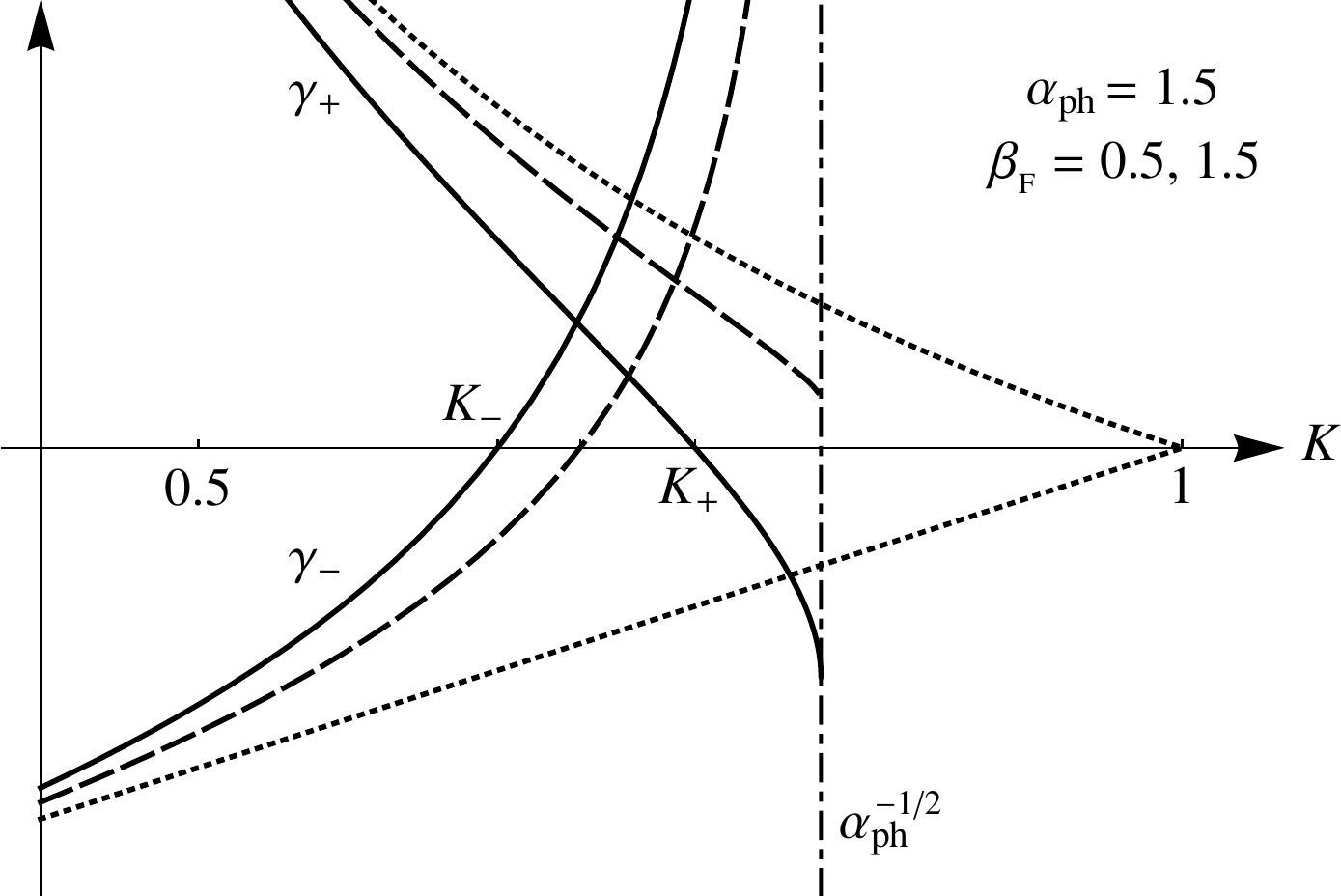}\\[9pt](b)}  \\  \caption{Weak tunneling ($\gamma_+$) and weak scattering ($\gamma_-$) exponents as functions of the Luttinger parameter $K<1$ for a weak (a) or strong (b) el-ph coupling, ($\alpha_{\rm ph}\lessgtr1$, respectively). The dotted lines represent $ \widetilde{\gamma}_\pm$ for the phononless LL,  the solid and dashed lines correspond to $\bF\equiv \vF/c=0.5$ and $\bF=1.5$, respectively. The latter is shown only for the strong coupling where for   $\bF>\big[\alpha_{\mathrm{ph}}(\alpha_{\mathrm{ph}}-1)\big]^{-1/2}$ the exponent $\gamma_+$ does not change sign within the stability region, $K<\alpha_\mathrm{ph}^{-1/2}$.}\label{gamma1}\end{figure*}
The gauge transform (\ref{gauge}) replaces $\lambda_0$ in Eq.~(\ref{WS}) with
\begin{align}\label{lambda}
 \lambda({t})&=\lambda_0\,\ee^{\ii\left[\theta_L(t)-\theta_R(t)\right]}\,,
 &
 \theta_{\eta}(t)&\equiv\theta_{\eta}(x\!=\!0,t)\,.
\end{align}
Integrating out all $x\!\ne\!0$ fields results in the quadratic in $\theta({t})$ action with the scattering term (\ref{lambda}) as in the phononless problem \cite{KaneFis:92a}. The  difference is that the  correlation function is governed by  the composite bosonic modes, Eq.~(\ref{U}), and its retarded Fourier component is
\begin{align}\label{u}
\langle\theta_{\eta}(-\omega)\,\theta_{\eta'}(\omega)\rangle^{r} =\ii\int \!\frac{\dd q}{2\pi}  \,U_{\eta\eta'}({\omega; q})\equiv -\frac{\ii\pi\gamma_{\eta\eta'}}{\omega_+}\,,
\end{align}
where the correlation matrix $\gamma_{\eta\eta'}$ is found from  the straightforward integration above as follows:
\begin{equation}\label{gamma}
\begin{aligned}
\gamma_{_\mathrm{LL}}=\gamma_{_\mathrm{RR}}&= \widetilde{\alpha} (K_{\rm fast})\,\xi_++ \widetilde{\alpha} (K_{\rm slow})\, \xi_-\,, \\
\gamma_{_\mathrm{LR}}=\gamma_{_\mathrm{LR}}&=  \alpha(K_{\rm fast})\,\xi_+ +\alpha(K_{\rm slow})\,\xi_-\,.
\end{aligned}
\end{equation}
Here $\xi_{\pm}=\frac{1}{2}\big[1\pm\big({v^2-c^2}\big) \big({v_+^2-v_-^2}\big)^{-1}\big]$ are the weight functions of the two modes and
\begin{align}\label{alpha}
\widetilde{\alpha}(K)&=\tfrac12(K^{-1}+K)-1,& \alpha(K)&=\tfrac12(K^{-1}-K ).
\end{align}
In the absence of phonons $\xi _+\!=\!1$ and $\xi _-\!=\!0$ and the fast mode contribution reproduces the results of \cite{KaneFis:92a}.

\begin{subequations}\label{RG}%
The only change in the $x\!=\!0 $ action which is due to the el-ph coupling is the appearance of the dimensionless
 matrix $\gamma_{\eta\eta'}$ in Eq.~(\ref{u}). Thus the RG analysis for the WS case is similar to that in \cite{KaneFis:92a} but  the RG equation for the backscattering amplitude $\lambda$  acquires a different prefactor  defined by Eq.~(\ref{gamma}), $\gamma_-=\gamma_{_\mathrm{LL}}-\gamma_{_\mathrm{LR}}$:
 \begin{align}\label{rgl}
\partial_{l}\,\lambda(E)&=\left\{\begin{array}{rr}
                           (1-K)\lambda(E)\,, & E > \omega_{_{\rm D}} \\
                           -\gamma_-\,\lambda(E)\,, & E < \omega_{_{\rm D}}
                         \end{array}\right.\,.
\end{align}
Here $E$ is a running cutoff and  $l\equiv\ln E_0/E$ with $E_0\sim\varepsilon_{_{\rm F}}$ being the bandwidth.

The WL case can be treated with the same  $x\!=\!0 $ action defined by Eq.~(\ref{u}) (rather than going to the dual $\phi $-action \cite{KaneFis:92a} which is inconvenient for our mixed representation \cite{noteGaYL1}). The gauge transform (\ref{gauge}) changes the tunneling action (\ref{WL}) by replacing $\psi_a \to \sum_\eta \psi_{a\eta}\ee^{\ii \eta \theta_a^-}$ and $t_0\to t_0\ee^{\ii({\theta_2^+-\theta_1^+})}$, where $\theta_a^\pm= \frac{1}{2} \big( \theta_{a\mathrm{R}}-\theta_{a\mathrm{L}}\big) $. The RG analysis of the tunneling term yields  the RG equation
\begin{align}\label{rgt}
\partial_{l}\,t(D)&=\left\{\begin{array}{rr}
                           (1-{K}^{-1})\,t(E)\,, & E > \omega_{_{\rm D}} \\[3mm]
                       -    \gamma_+\,t(E)\,, & E < \omega_{_{\rm D}}
                         \end{array}\right.\,,
\end{align}
\end{subequations}
dual to Eq.~(\ref{rgl}), with $\gamma_+=\gamma_{_\mathrm{LL}}+\gamma_{_\mathrm{LR}}$.

The RG equations (\ref{RG}) should be solved with the initial conditions $\lambda(E\!=\!E_0)=\lambda_0$ and $t(E\!=\!E_0)=t_0$. The solutions for $\varepsilon <\omega_{_{\rm D}}$ are given by
\begin{align}\label{RGs}
\frac{\lambda(\varepsilon)}{\lambda_0}&=\left(\frac{\omega_{_{\rm D}}}{E_0}\right)^{\!\!\widetilde\gamma_-}\!\!\! \left(\frac{\varepsilon} {\omega_{_{\rm D}}} \right)^{\!\!\gamma_-}\!\!\!,&
\frac{t(\varepsilon)}{t_0}&=\left(\frac{\omega_{_{\rm D}}}{E_0}\right)^{\!\!\widetilde\gamma_+}\!\!\! \left(\frac{\varepsilon} {\omega_{_{\rm D}}} \right)^{\!\!\gamma_+}\!\!\!.
\end{align}
The exponents $\gamma_\pm$  can be rewritten (with $\beta\equiv v/c$) as
\begin{subequations}\label{gk}%
\begin{align}\label{gs}
&\gamma_{\sigma}= {\kappa_\sigma}{K}^{-\sigma}-1\,, \qquad\sigma =\pm1\,,\\\label{kappa}
&\begin{aligned}
    \kappa_\sigma&=\bigg\{\left[1\!-\! \alpha\delta_{\sigma,-1}\right]\bigg[1 +\frac{\alpha} {\big(\beta^\sigma+\sqrt{1\!-\!\alpha}\,\big)^2 }
    \bigg]
    \bigg\}^{\!-\frac{1 }{2} }\!,
\end{aligned}
\end{align}
\end{subequations}
while $\widetilde\gamma_\pm$ are obtained with $\kappa_{\pm}\!=\!1$ in Eq.~(\ref{gs}).
In the absence of the el-ph coupling we have $\gamma_\sigma =\widetilde \gamma_\sigma$ and Eqs.~(\ref{RGs}) reduce to the standard ones, $\lambda({\varepsilon })\sim({\varepsilon /E_0})^{K-1 }$ and $t({\varepsilon })\sim ({\varepsilon /E_0})^{K^{-1}-1}$, i.e.\ the backscattering amplitude  increases (the WS case) and the tunneling amplitude decreases (the WL case) with $\varepsilon \to0$ for any $K<1$. This shows that embedding an arbitrary scatterer results in  the  LL becoming (at $T=0$)   an ideal insulator.

The  el-ph coupling, however weak,  changes drastically  the above conclusion leading to the possibility of a metal-insulator transition with changing $K$.

Indeed, it follows from Eqs.~(\ref{RG})--(\ref{gk}) that RG flows can change directions depending on the values of $K$, $\alpha_\mathrm{ph}$ and $\bF\equiv\vF/c$. It is easy to verify that $\kappa_+(K)\leq 1$ while $\kappa_-(K)\geq 1$ which results in $\gamma_\pm$ changing sign (see Fig.~1) when the interaction strength $K$ equals $K_\pm$, with
\begin{align}\label{K}
\gamma_+(K_+)&=0\,,&\gamma_-(K_-)&=0\,;& K_-<K_+<1\,.
\end{align}
Then for $K_+<K<1$ both $\gamma_\pm$ change sign so that the WL amplitude $t({\varepsilon })$ increases while the WS amplitude $\lambda({\varepsilon })$ decreases with $\varepsilon \to 0$ in Eqs.~(\ref{RGs}), indicating \cite{noteGaYL2} that adding a single scatterer does not change the metallic nature of the pure LL (where the dimensionless conductance $g=1$) -- in contrast to the phononless case.

Decreasing $K$ (i.e.\ increasing the el-el interaction), we enter the region $K_-<K<K_+$, when the WL amplitude still decreases -- as in the phononless case, but also the WS amplitude decreases -- opposite to the phononless case. This indicates that a strong scatterer (i.e.\ weak link) results in the insulating behavior (the $g=0$ fixed point) while a weak scatterer leaves the LL in the metallic phase (the $g=1$ fixed point). Thus there should exist an intermediate unstable fixed point with a finite $g$, separating the metallic and insulating regimes.

Finally, for a strong enough el-el interaction, $K<K_-$, the RG flows  in Eqs.~(\ref{RGs}) remain for both the WL and WS amplitudes qualitatively the same as for the phononless case, so that any scatterer makes the LL insulating.

Explicitly, we find $K\pm$ for  $\alpha_\mathrm{ph}\ll1$ as follows:
\begin{align*}
   1- K_\sigma&\approx \frac{\alpha_\mathrm{ph}\sigma}{2}\Bigl[\delta_{\sigma,-1} -\frac{1}{(\bF +1)^2}\Bigr]\,,&\bF&\equiv\frac{\vF}{c}\,.
\end{align*}
This means that  in this limit there exist all the three regimes described above, see Fig.~1(a).

When the el-ph coupling is not weak (which is the case for carbon nanotubes  \cite{DeMartinoEgger:03,*sapmaz}) our considerations are bound by the stability requirement, $K\leq\alpha^{-1/2}_{\rm ph}$ so that for $\alpha_{\mathrm{ph}}>1 $ the Luttinger parameter is confined to the region $K\leq \alpha^{-1/2}_{\rm ph}$.  In this case there is an essential dependence on $\bF$.  It is easy to see from Eqs.~(\ref{gk}) and (\ref{K}) that $K_+>\alpha^{-1/2}_{\rm ph}$ for $\bF^2 >\big[\alpha_{\mathrm{ph}}(\alpha_{\mathrm{ph}}-1)\big]^{-1}$  so that for such $\bF$ the purely metallic regime (with both $\gamma_\pm$ changing sign) is no longer accessible, as illustrated by the dashed lines in Fig.~1b.  On the phase diagram ({Fig.~2}) all the three regimes described above exist only for $\alpha_{\mathrm{ph}}<\alpha_{\mathrm{ph}}^*\equiv \frac{1}{2}+\frac{1}{2}\sqrt{1+4/\bF^2}  $.

There are other properties of the LL strongly affected by the el-ph coupling.  First, $\gamma_+$ is also the edge exponent of the TDoS at the boundary (edge) of the wire, $\nu_\mathrm {edge}\sim\varepsilon^{\gamma_+}$, the TDoS also experiences, depending on the parameters, a transition from vanishing  to divergent with $\varepsilon \to 0$. Then,  transport properties of the LL with resonant or antiresonant impurity \cite{KaneFis:92b,*PG:03,*FurMatv:02,*NG,*LYY,*GB:10}, or of the disordered LL    \cite{GiamarchiSchulz}, or of the LL out of equilibrium \cite{GGM:08,*GGM:09,*GGM:10}  also experience qualitative changes with allowing for the el-ph interaction, as we will show elsewhere.

\begin{figure}  \includegraphics[width=.9\columnwidth]{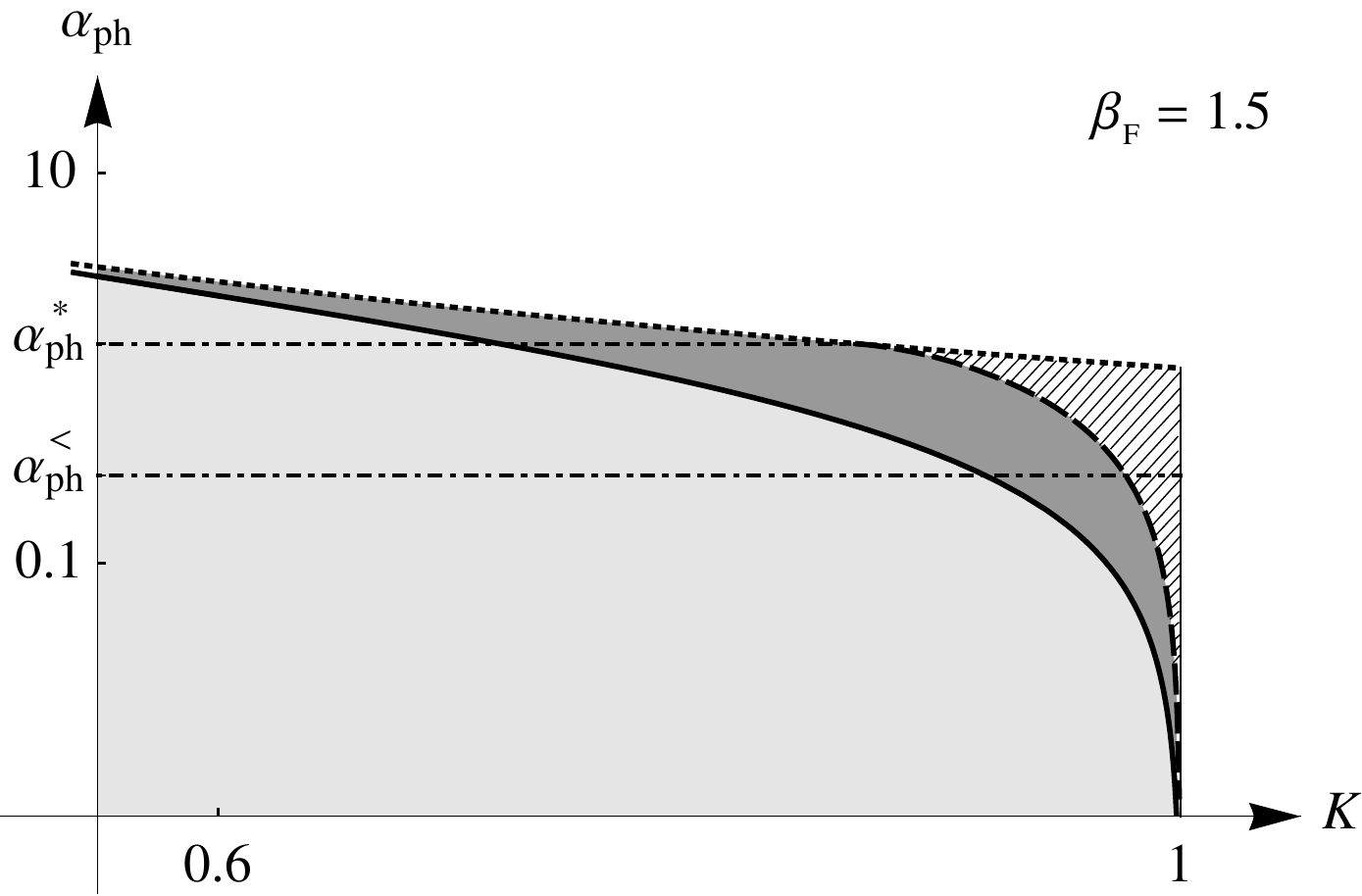}  \caption{Phase diagram: the dotted line is the boundary of the stability area, $\alpha_\mathrm{ph}=K^{-2}$, the solid and dashed lines correspond to $K_\mp$, respectively. The light-grey area represents  the insulator (both $\gamma_\pm$ have the same sign as in the absence of the el-ph coupling). In the dark-grey  area (where only $\gamma_-$ changes sign) there exists a line of unstable fixed points  separating the metallic state for a weak scatterer from the insulating phase for a strong one. The hatched area (reachable only for $\alpha_\mathrm{ph}<\alpha_\mathrm{ph}^*$, as illustrated by some $\alpha_{\mathrm{ph}}^< $) corresponds to the purely metallic phase  where both $\gamma_\pm$ change signs.}\end{figure}

To summarize, we have shown that the el-ph coupling  qualitatively changes the phase diagram of the LL with a single impurity.
The change is not reducible to a redefinition of the Luttinger parameter $K$: the existence of slow and fast polaron modes with different weights in different regimes results in different RG flows for a weak scatter and a weak (tunneling) link.  The resulting phase diagram (Fig.~2)  has, depending on the parameters of the problem,  regimes corresponding to purely metallic or purely insulating behavior and an intermediate regime with two stable fixed points (ideal metal and insulator) and one unstable, finite-conductance fixed point.

\begin{acknowledgements}
    This work was supported by the EPSRC Grant T23725/01. I.V.Y.\ and I.V.L.\  gratefully acknowledge kind hospitality at the Abdus Salam ICTP.
\end{acknowledgements}


\begin{thebibliography}{37}
\bibitem [{not(){\natexlab{a}}}]{noteGaYL1}%
  \BibitemOpen
  \href@noop {} {}   \bibinfo {note} {After the gauge
  transformation (\ref{gauge}) the remaining fermionic part can be bosonized
  with $\psi\sim\ee^{i\phi }$. This would allow one to rewrite the action in the standard, fully bosonized form in terms of $\Theta\equiv\theta+\phi $ and $\phi $. However, we will use the mixed fermion-boson
  representation.}\BibitemShut {Stop}%
\end{thebibliography}

%

\end{document}